\documentclass[prc,twocolumn,showpacs,showkeys,nofootinbib,superscriptaddress]{revtex4-1}
\usepackage{graphicx}
\usepackage{dcolumn}
\usepackage{epsfig}
\usepackage{bm}

\usepackage{amsfonts}
\usepackage{amsmath}
\usepackage{amssymb}
\usepackage[english]{babel}
\usepackage{graphicx}
\usepackage{epsfig}
\usepackage{bm}
\usepackage{verbatim} 
\usepackage[utf8]{inputenc}
\usepackage{booktabs}
\usepackage{multirow}
\usepackage[dvipsnames,table,xcdraw]{xcolor}
\usepackage{float}
\usepackage{blindtext}

\usepackage[breaklinks=true]{hyperref}
\hypersetup{colorlinks=true, linkcolor=blue, urlcolor=blue, citecolor=blue}

\begin{document}
	
	\title{A Semi-Empirical Formula for Two-Neutrino Double-Beta Decay}
	
	\author{O. Ni\c{t}escu}
	\affiliation{Faculty of Mathematics, Physics and Informatics, Comenius University in Bratislava, 842 48 Bratislava, Slovakia}
	\affiliation{ International Centre for Advanced Training and Research in Physics, P.O. Box MG12, 077125 M\u{a}gurele, Romania}
	\affiliation{“Horia Hulubei” National Institute of Physics and Nuclear Engineering, 30 Reactorului, POB MG-6, RO-077125 Bucharest-M\u{a}gurele, Romania}
	
	\author{F. \v{S}imkovic}
	\email{fedor.simkovic@fmph.uniba.sk}
	\affiliation{Faculty of Mathematics, Physics and Informatics, Comenius University in Bratislava, 842 48 Bratislava, Slovakia}
	\affiliation{Institute of Experimental and Applied Physics, Czech Technical University in Prague,
		110 00 Prague, Czech Republic}
	
	\begin{abstract}
		We propose a semi-empirical formula (SEF) for calculating the nuclear matrix elements (NMEs) for two-neutrino double-beta decay. The SEF's dependence on the proton and neutron numbers, the pairing, isospin, and deformation properties of the initial and final nuclei is inspired by the insights offered by nuclear many-body methods and trends in experimental data. Compared with the previous phenomenological and nuclear models, the SEF yields the best agreement with the experimental NMEs. Its stability and predictive power are cross-validated, and predictions are provided for nuclear systems of experimental interest.
	\end{abstract}
	
	\maketitle

	\section{Introduction}
	
	In 1935, about a year after the Fermi weak interaction theory was introduced \cite{Fermi-ZP1934}, Goeppert-Mayer considered, at the suggestion of Wigner, the two-neutrino double-beta decay ($2\nu\beta\beta$-decay) process \cite{Goeppert-Mayer-PR1935}:
	\begin{equation}
		(A, Z_i) \rightarrow (A, Z_i+2) + e^- + e^- + \overline{\nu}_e  + \overline{\nu}_e.
	\end{equation}
	In this process, two neutrons in the initial nucleus, $(A, Z_i)$, where $A$ and $Z_i$ are the mass and atomic numbers, respectively, are converted into two protons while emitting two electrons and two electron antineutrinos. The half-life of $2\nu\beta\beta$-decay was estimated to be $10^{17}$ years, assuming a $Q$-value of about 10 MeV \cite{Goeppert-Mayer-PR1935}. In 1939, after the theory of Majorana neutrinos was introduced \cite{Majorana-INC1937}, Furry proposed the concept of neutrinoless double-beta decay ($0\nu\beta\beta$-decay) \cite{Furry-PR1939},
	\begin{equation}
		(A, Z_i) \rightarrow (A, Z_i+2) + e^- + e^-,
	\end{equation}
	involving two subsequent $\beta$ decays of neutrons connected via the exchange of virtual neutrinos at that time \cite{Racah-INC1937}. Today, other mechanisms that could trigger $0\nu\beta\beta$-decay are explored \cite{Vergados-IJMPE2016,Cirigliano-JHEP2018,Graf-PRD2018}.  There are other modes of double-beta decay transforming the $(A,Z_i)$ nucleus into the $(A,Z_i-2)$ nucleus \cite{Doi-PTP1992,Doi-PTP1993}: the double-positron emitting ($2\nu/0\nu\beta^+\beta^+$) mode, the atomic electron capture with coincident positron emission ($2\nu/0\nu\textrm{EC}\beta^+$) mode, and the double electron capture ($2\nu/0\nu\textrm{ECEC}$) mode. These processes are less favored due to smaller $Q$-value, small overlap of the bound electron wave function with the nucleus, and Coulomb repulsion on positrons.
	
	The primary focus is on studying $0\nu\beta\beta$-decay, which is not allowed by the  Standard Model (SM) of particle physics. Thus, it can provide insight into neutrinos' Majorana nature, masses, and CP properties by violating the total lepton number \cite{Schechter-PRD1982,Sujkowski-PRC2004,Pascoli-NPB2006,Bilenky-IJMPA2015,Vergados-IJMPE2016,Girardi-NPB2016,Simkovic-PU2021}. To date, the $0\nu\beta\beta$-decay has not been observed. The best lower limit on the half-life of this process is approximately $T_{1/2}^{0\nu}\gtrsim 10^{26}$ years for $^{76}$Ge \cite{GERDA-PRL2020} and $^{136}$Xe \cite{KAMLAND-ZEN-PRL2023,KamLAND-Zen-arXiv2024}. The discovery of $0\nu\beta\beta$-decay would have far-reaching implications for particle physics and fundamental symmetries.
	
	The first direct observation of $2\nu\beta\beta$-decay, allowed by the  SM, was achieved for $^{82}$Se in 1987 \cite{Elliott-PRL1987,Moe-ARNPS2014}. Nowadays, the $2\nu\beta\beta$-decay has been detected in direct counter experiments in nine different nuclei, including decays into two excited states. In addition to laboratory experiments, geochemical and radiochemical observations of $2\nu\beta\beta$-decay transitions have been recorded \cite{Barabash-U2020} (and references therein). There are also positive indications of the $2\nu\textrm{ECEC}$ mode for $^{130}$Ba and $^{132}$Ba from geochemical measurements \cite{Barabash-PAN1996, Meshik-PRC2001, Pujol-GCA2009}, as well as for $^{78}$Kr \cite{Gavrilyuk-PRC2013,Ratkevich-PRC2017}. Recently, the first direct observation of the $2\nu\textrm{ECEC}$ process in $^{124}$Xe was reported \cite{XENON1T-N2019, XENON1T-PRC2022,LUX-ZEPLIN-JPG2024,PandaX-arXiv2024}. The measurement of $2\nu\beta\beta$-decay provides valuable insights into nuclear structure physics, which can be used in $0\nu\beta\beta$-decay calculations and the search for new physics beyond the SM \cite{Bossio-JPG2024}.
	
	Despite the recent advancement in measuring the $2\nu\beta\beta$-decay and ongoing efforts to develop a robust theoretical description, mismatches between theory and experiment persist. One issue is the significant spread observed in experimental $2\nu\beta\beta$-decay nuclear matrix elements (NMEs), which cannot be uniformly explained within existing nuclear models without fine-tuning various parameters on a case-by-case basis. In this paper, we propose a semi-empirical formula (SEF) for calculating the $2\nu\beta\beta$-decay NMEs. The paper is organized as follows: Section~\ref{sec:CurrentState} reviews the current state of NME calculations, covering existing nuclear structure models, the single-state dominance hypothesis, and previous phenomenological models. In Section~\ref{sec:SEF}, we present the proposed SEF which, inspired by trends in the nuclear data and the findings of nuclear many-body methods, includes the proton and neutron numbers, the pairing, isospin, and deformation properties of the initial and final nuclei. Section~\ref{sec:RD} details the determination of fit parameters and demonstrates, through comparisons with prior phenomenological and nuclear models, that the proposed SEF achieves the best agreement with experimental data. This section also presents cross-validation results and predictions for nuclear systems of interest. Section~\ref{sec:Conclusions} provides concluding remarks.            
	
	\section{Current state}
	\label{sec:CurrentState}
	
	The inverse $2\nu\beta\beta$-decay half-life is commonly presented as 
	\begin{equation}
		\left( T^{2\nu}_{1/2}\right)^{-1} = \left| M^{2\nu} \right|^2 ~ G^{2\nu},
	\end{equation}
	where $G^{2\nu}$ is the phase-space factor (PSF), and
	\begin{equation}
		M^{2\nu} = g^2_A  M^{2\nu}_{GT}  - g_V^2 M^{2\nu}_F,
	\end{equation}
	is the NME governing the transition. $g_V=1$ and $g_A$ are the vector and axial-vector coupling constants, respectively. The latter is usually model-dependent and remains an open question in nuclear weak interaction processes \cite{Suhonen-FP2017}. The impulse approximation for nucleon current and only $s_{1/2}$ wave states of emitted electrons are considered. The Fermi $M_F^{2\nu}$ and Gamow-Teller (GT) $M_{GT}^{2\nu}$  matrix elements are governed by the Fermi and GT operators, which are generators of isospin SU(2) and spin-isospin SU(4) symmetries, respectively. The isospin is known to be a good approximation in nuclei. Thus, it is assumed that $M^{2\nu}_F$ is negligibly small, and the main contribution is given by $M^{2\nu}_{GT}$, which takes the form
	\begin{eqnarray}
		M^{2\nu}_{GT}  = m_e \sum_n \frac{M_n}{E_n - (E_i + E_f)/2},
	\end{eqnarray}
	with 
	\begin{eqnarray}
		M_n =
		\langle 0^+_f\| \sum_{j}\tau^+_{j} \sigma_{j} \| 1^+_{n}\rangle
		\langle 1^+_n \| \sum_{k}\tau^+_{k} \sigma_{k} \| 0^+_{i} \rangle.
	\end{eqnarray}
	Here, $|0^+_i\rangle$ ($|0^+_f\rangle$) is the ground state of the initial (final) even-even nucleus with energy $E_i$ ($E_f$), and the summations run over all $|1^+_n\rangle$ states the intermediate odd-odd nucleus with energies $E_n$ and over all $j,k $ nucleons inside the nucleus.

	The phase-space part of the decay rate can be accurately computed with a relativistic description of the emitted (captured) electrons in a realistic potential of the final (initial) atomic system. The most precise PSFs computed within the self-consistent Dirac-Hartree-Fock-Slater method, including radiative and atomic exchange corrections \cite{Nitescu-PRC2023,Nitescu-PRC2024}, are presented in Table~\ref{tab:syst}. On the other hand, the computation of the NMEs is a challenging and long-standing problem in this field. Due to the growing interest in detecting neutrinoless modes, there have been significant reductions in the uncertainties of half-lives for two-neutrino modes (see column 11 in Table~\ref{tab:syst}) and consequently of the experimental NMEs, i.e., $M^{2\nu-\textrm{exp}}=(T^{2\nu{\rm -exp}}_{1/2}G^{2\nu})^{-1/2}$ (see Table~\ref{tab:modelComp} and Table~\ref{tab:phComp}). The spread and structure of the experimental NMEs are still poorly understood. A model that aligns with the current experimental NMEs could provide realistic recommendations for future experiments across the nuclear chart.
	
	\begin{table*}
		\squeezetable
		\caption{The decay properties for various $2\nu\beta\beta$-decays (right arrow) and $2\nu\textrm{ECEC}$ processes (left arrow).  $T_{f}$ represents the total isospin of the final nucleus. $\Delta_p$ and $\Delta_n$ denote experimental proton and neutron pairing gaps in units of MeV, respectively. The quadrupole deformation parameters, $\beta^{\rm Z, Z+2}$, deduced from the corresponding B(E2) transitions, are taken from Brookhaven Nuclear Database \cite{NNDC}. $G^{2\nu}$ in $\textrm{yr}^{-1}$ is the most precise PSF computed in this work for $2\nu\beta\beta$ cases and taken from \cite{Nitescu-U2024} ($^{78}$Kr, $^{106}$Cd, $^{130}$Ba, and $^{132}$Ba) and \cite{Nitescu-JPG2024} ($^{124}$Xe) for the $2\nu\textrm{ECEC}$ cases. $T_{1/2}^{2\nu{\rm -ph}}$ and $T_{1/2}^{2\nu{\rm -exp}}$ are the half-lives predicted by the SEF and the experimental half-lives with the smallest uncertainties, respectively. The $^{78}$Kr data are for double electron capture from $K$ orbital. For $^{128}$Te, the experimental half-life is obtained using the ratio $T^{2\nu{\rm -exp}}_{1/2}(^{130}\textrm{Te})/T^{2\nu{\rm -exp}}_{1/2}(^{128}\textrm{Te})=\left(3.52\pm0.11\right)\times10^{-4}$ \cite{Bernatowicz-PRC1993}. The transitions under the "Fitted" label are used as input for the SEF, while those under the "Predicted" label are its predictions. \label{tab:syst}}
		\centering
		\begin{ruledtabular}
			\begin{tabular}{lcccccccccr}
				Nuclear transition &   $T_{f}$  &  $\Delta^{\rm Z}_p $ & $\Delta^{\rm Z}_n  $ & $\Delta^{\rm Z+2}_p$ & $\Delta^{\rm Z+2}_n$  & $\beta^{Z}$ & $\beta^{Z+2}$  &$G^{2\nu}$ &  $T^{2\nu{\rm -ph}}_{1/2}$& $T^{2\nu{\rm -exp}}_{1/2}$  \\\hline
				
				Fitted&&&&&&&&&&\\ 
				\hline
				${^{48}_{20}{\rm Ca}_{ 28}}\rightarrow {^{48}_{22}{\rm Ti}_{26}}$ & 2 & 2.179  & 1.688 & 1.896 & 1.564 &   0.107  & 0.262  & $1.58\times10^{-17}$ &$1.2\times10^{20}$ &$6.4^{+1.3}_{-1.1}\times 10^{19}$\cite{Arnold-PRD2016-48Ca}\\
				
				${^{76}_{32}{\rm Ge}_{44}}\rightarrow {^{76}_{34}{\rm Se}_{42}}$ & 4 &
				1.561 & 1.535 & 1.751 & 1.710 &  0.188  & 0.219  & $5.07\times10^{-20}$ &$1.8\times10^{21}$ &$2.022^{+0.042}_{-0.042}\times 10^{21}$\cite{Agostini-PRL2023-76Ge}\\
				
				${^{82}_{34}{\rm Se}_{48}}\rightarrow {^{82}_{36}{\rm Kr}_{46}}$ & 5 & 1.401 & 1.544 & 1.734 & 1.644  &  0.194  & 0.203  & $1.68\times10^{-18}$ &$9.1\times10^{19}$ &$8.69^{+0.10}_{-0.07}\times 10^{19}$\cite{Azzolini-PRL2023-82Se}\\

				${^{96}_{40}{\rm Zr}_{56}}\rightarrow {^{96}_{42}{\rm Mo}_{54}}$ & 6 &
				1.539 & 0.846 & 1.528 & 1.034 &  0.060  & 0.172  & $7.24\times10^{-18}$ &$3.4\times10^{19}$ &$2.35^{+0.21}_{-0.21}\times 10^{19}$\cite{Argyriades-NPA2010-96Zr}\\
				
				${^{100}_{42}{\rm Mo}_{58}}\rightarrow {^{100}_{44}{\rm Ru}_{56}}$ & 6 &
				1.612 & 1.358 & 1.548 & 1.296 &   0.231  & 0.215  & $3.47\times10^{-18}$ &$7.2\times10^{18}$ &$7.07^{+0.11}_{-0.11}\times 10^{18}$\cite{Augier-PRL2023-100Mo}\\
				
				${^{116}_{48}{\rm Cd}_{68}}\rightarrow {^{116}_{50}{\rm Sn}_{66}}$ & 8 &
				1.493 & 1.377 &  1.763  & 1.204  &  0.135  & 0.112  & $2.91\times10^{-18}$ &$2.3\times10^{19}$ &$2.63^{+0.11}_{-0.12}\times 10^{19}$\cite{Barabash-PRD2018-116Cd}\\
				
				${^{130}_{52}{\rm Te}_{78}}\rightarrow {^{130}_{54}{\rm Xe}_{76}}$ & 11&
				1.104 & 1.180 & 1.307 & 1.248 &   0.118  & 0.128 & $1.62\times10^{-18}$ &$7.5\times10^{20}$ &$8.76^{+0.17}_{-0.18}\times 10^{20}$\cite{Adams-PRL2021-130Te,Adams-PRL2023-130TeErratum}\\
				
				${^{136}_{54}{\rm Xe}_{82}}\rightarrow {^{136}_{56}{\rm Ba}_{80}}$ & 12 &
				1.009 & { 1.436}  & 1.265 & 1.031 &  0.091  & 0.125 & $1.52\times10^{-18}$ &$2.5\times10^{21}$ &$2.17^{+0.06}_{-0.06}\times 10^{21}$\cite{Albert-PRC2014-136Xe}\\
				
				${^{150}_{60}{\rm Nd}_{90}}\rightarrow {^{150}_{62}{\rm Sm}_{88}}$ & 13 &
				1.224 & 1.046 & 1.444 & 1.193 & 0.285   &  0.193 & $3.85\times10^{-17}$ &$2.5\times10^{19}$ &$9.3^{+0.7}_{-0.6}\times 10^{18}$\cite{Arnold-PRD2016-150Nd}\\
				
				${^{238}_{92}{\rm U}_{ 146}}\rightarrow {^{238}_{94}{\rm Pu}_{144}}$ & 25 &
				0.813 & 0.606 & 0.662 & 0.589 &  0.289 & 0.285 & $1.65\times10^{-19}$ &$9.1\times10^{21}$ &$2.0^{+0.6}_{-0.6}\times 10^{21}$\cite{Turkevich-PRL1991-238U}\\

				\hline
				Predicted&&&&&&&&&&\\
				\hline
				${^{110}_{46}{\rm Pd}_{64}}\rightarrow {^{110}_{48}{\rm Cd}_{62}}$ & 7 &
				1.422 & 1.442 & 1.479 & 1.362 &  0.257  & 0.172& $1.46\times10^{-19}$ &$3.1\times10^{20}$ &$> 10^{18}$\cite{Winter-PR1952-110PdLimit}\\
				
				${^{124}_{ 50}{\rm Sn}_{74}}\rightarrow {^{124}_{52}{\rm Te}_{72}}$ & 10 & 1.671 & 1.314 & 1.248 & 1.343 &  0.095 & 0.170 & $6.00\times10^{-19}$ &$1.9\times10^{21}$ &--\\
				
				${^{128}_{52}{\rm Te}_{76}}\rightarrow {^{128}_{54}{\rm Xe}_{74}}$ &10 &1.129 & 1.280 & 1.319 & 1.265 & 0.135 & 0.202& $3.00\times10^{-22}$ &$1.5\times10^{24}$ &$2.49^{+0.09}_{-0.09}\times 10^{24}$\\
				
				${^{134}_{54}{\rm Xe}_{ 80}}\rightarrow {^{134}_{56}{\rm Ba}_{78}}$ & 11 &
				1.134 & 1.029  & 1.329 & 1.176 &  0.115 & 0.163 & $2.25\times10^{-23}$ &$4.0\times10^{25}$ &$>2.8\times10^{22}$\cite{Yan-PRL2024-134XeLimit}\\

				\hline
				Fitted&&&&&&&&&&\\ 
				\hline
				${^{78}_{34}{\rm Se}_{44}} \leftarrow  {^{78}_{36}{\rm Kr}_{42}}$ & 5 &
				1.442 & 1.637 & 1.577 & 1.691 &  0.270  &0.256 & $5.20\times10^{-22}$&$1.7\times10^{23}$ &$1.9^{+1.3}_{-0.8}\times 10^{22}$\cite{Ratkevich-PRC2017-78Kr}\\
				
				${^{124}_{52}{\rm Te}_{72}} \leftarrow  {^{124}_{54}{\rm Xe}_{70}}$ & 10 &
				1.249 & 1.344 & 1.357 & 1.339 &  0.170  & 0.223  & $1.83\times10^{-20}$ &$2.0\times10^{22}$ &$1.10^{+0.22}_{-0.22}\times 10^{22}$\cite{Aprile-PRC2022-124Xe}\\

				\hline
				Predicted&&&&&&&&&&\\
				\hline
				${^{106}_{46}{\rm Pd}_{60}}\leftarrow {^{106}_{48}{\rm Cd}_{58}}$ &7 &
				1.472 & 1.401 & 1.460 & 1.338 &  0.162 &0.168 & $5.54\times10^{-21}$ &$2.1\times10^{21}$ &$>4.7\times 10^{20}$\cite{Belli-U2020-116PdLimit}\\
				
				${^{130}_{54}{\rm Xe}_{76}}\leftarrow {^{130}_{56}{\rm Ba}_{74}}$  &11 &	1.308 & 1.248  & 1.351 & 1.298 & 0.128 & 0.215& $1.57\times10^{-20}$ &$6.0\times10^{22}$ &$2.2^{+0.5}_{-0.5}\times 10^{21}$\cite{Meshik-PRC2001-130BaLimit}\\
				
				${^{132}_{54}{\rm Xe}_{78}}\leftarrow {^{132}_{56}{\rm Ba}_{76}}$ &   12   & 1.240 & 1.181  & 1.390 & 1.236 & 0.141 & 0.185 & $4.09\times10^{-23}$ &$1.0\times10^{26}$ &$>2.2\times 10^{21}$\cite{Barabash-PAN1996-132BaLimit}\\
			\end{tabular}
		\end{ruledtabular}
	\end{table*}

	Currently, the predictions of the $2\nu\beta\beta$-decay NMEs are based on various nuclear structure models, such as the proton-neutron quasiparticle random phase approximation (pn-QRPA) and its variants \cite{Vogel-PRL1986,Civitarese-PLB1987,Aunola-NPA1996a,Aunola-NPA1996b,Suhonen-PR1998,Stoica-NPA2001,Stoica-TEPJA2003,Simkovic-NPA2004,Alvarez-PRC2004,Rodin-NPA2006,Suhonen-JPG2012,Simkovic-PRC2013,Pirinen-PRC2015,Simkovic-PRC2018}, the nuclear shell model (NSM) \cite{Caurier-PRL1996,Caurier-RMP2005,Horoi-PRC2007,Caurier-PLB2012,Neacsu-PRC2015,Brown-PRC2015,Horoi-PRC2016,Kostensalo-PLB2020,Patel-NPA2024}, the interacting boson model (IBM) \cite{Barea-PRC2013,Barea-PRC2015,Nomura-PRC2022,Nomura-PRC2024}, the projected Hartree-Fock-Bogoliubov (PHFB) method \cite{Rath-FP2019}, Fermi surface quasi-particle (FSQP) model \cite{Ejiri-JPSJ2005,Ejiri-JPG2017},  effective theory (ET) \cite{CoelloPerez-PRC2018}, and others \cite{Kotila-JPG2010,Popara-PRC2022}, as well as phenomenological models \cite{Ren-PRC2014,Rajan-IJP2018,Pritychenko-NPA2023}. Calculating the $2\nu\beta\beta$-decay NMEs in nuclear theory is challenging due to the complex structure of open-shell medium and heavy nuclei, and the need to describe a complete set of intermediate nucleus states.
	
	Two approaches have been used for a long time: the pn-QRPA and the NSM. Although limited to low-lying excitations, the NSM accounts for all possible correlations within the valence space. It should be noted that higher-lying excitations can be accessed through the use of Lanczos strength function method \cite{Caurier-RMP2005}.   To match the $2\nu\beta\beta$-decay data, the Gamow–Teller operator is adjusted (quenched) to fit the results of single $\beta$ decays or charge exchange reactions, and an unquenched $g_A$ is assumed \cite{Caurier-PRL1996,Caurier-PLB2012}. The pn-QRPA involves orbitals far from the Fermi surface, considering high-lying excited states (up to 20-30 MeV), but it includes fewer correlations. It was found that $M^{2\nu}_{GT}$ and $M^{2\nu}_F$ depend sensitively on the isoscalar and isovector particle-particle interactions of nuclear Hamiltonian, respectively \cite{Stefanik-PRC2015,Simkovic-PRC2018}. The IBM simplifies the low-lying states of the nucleus into $L = 0$ ($s$ boson) or $L = 2$ ($d$ boson) states, focusing on $0^+$ and $2^+$ neutron pairs converting into two protons. The PHFB formalism obtains nuclear wave functions with good particle number and angular momentum by projecting on axially symmetric intrinsic HFB states. However, it limits the nuclear Hamiltonian to quadrupole interactions. Although not covered in this study, due to the challenge of obtaining systematic results, we note the recent progress toward \textit{ab initio} descriptions of $2\nu\beta\beta$-decay transitions. Notable examples include coupled-cluster calculations for $^{48}$Ca \cite{Novario-PRL2021} and hybrid calculations that combine chiral effective-field theory and the NSM for $^{48}$Ca, $^{76}$Ge, and $^{82}$Se \cite{Coraggio-PRC2024}.

	Phenomenological models aim to simplify the process description. For example, according to \cite{Abad-INCA1983,Abad-JPC1984}, $2\nu\beta\beta$-decays with a $1^+$ ground state of the intermediate nucleus are determined solely by two virtual $\beta$ decay transitions. The first connects the initial nucleus with the $1^+$ intermediate ground state, and the second connects this $1^+$ state with the final ground state. This assumption is the single-state dominance (SSD) hypothesis. One advantage of SSD is its independence from the nuclear structure model, as $M^{2\nu}$ can be determined from measured $\log ft$ values or charge-changing reactions. However, a recent study on electron energy distributions in $2\nu\beta\beta$-decay of $^{100}$Mo \cite{Augier-PRL2023-100Mo} suggests that transitions through higher-lying states of the intermediate nucleus also play a crucial role, with destructive interference occurring between these contributions to $M^{2\nu}$.

	To our knowledge, three other phenomenological models have been proposed for predicting the $2\nu\beta\beta$-decay half-lives or NMEs. One model \cite{Ren-PRC2014} is similar to the Geiger-Nuttall law used for $\alpha$ decay. The other two models \cite{Rajan-IJP2018,Pritychenko-NPA2023} suggest that half-lives or NMEs depend on parameters such as the Coulomb energy parameter ($\xi \approx Z A^{-1/3}$), the $Q$-value, and the quadrupole deformation parameter of the initial nucleus. However, it will be shown that these models could benefit from refinements to better capture the complexity of the problem.             
	
	\vspace{-0.4cm}
	
	\section{Semi-empirical formula for $2\nu\beta\beta$-decay}
	\label{sec:SEF}
	
	\begin{figure*}
		\includegraphics[width=1.5\columnwidth]{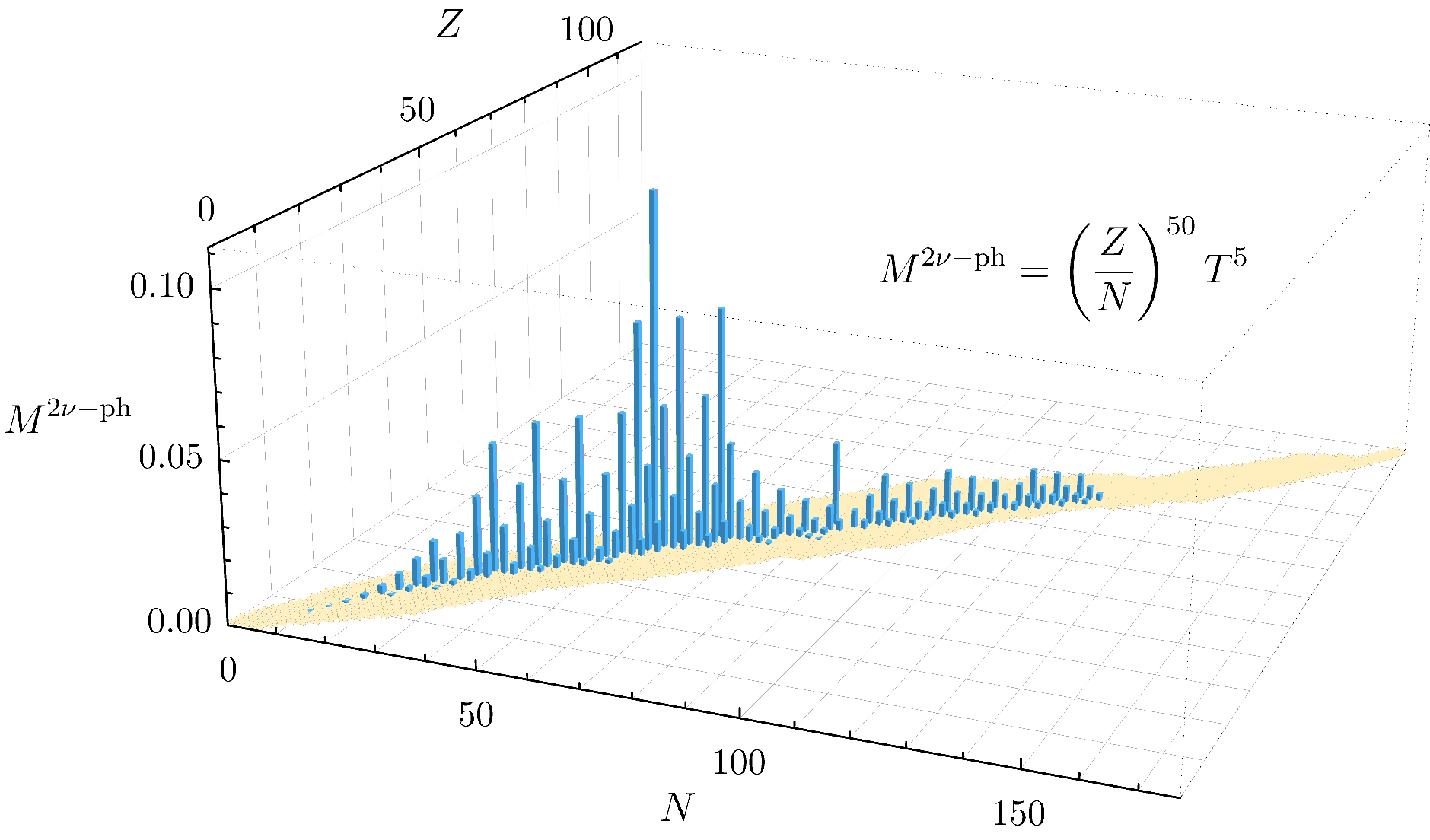}
		\caption{\label{fig:epsart} The phenomenological NMEs, obtained using a simplified model $M^{2\nu{-\textrm{ph}}}=(Z/N)^{50}~T^5$, are represented in the $(Z,N)$ space for stable even-even nuclei. The base of the representation, shown in orange, corresponds to the known nuclear chart.}
	\end{figure*} 
	
	This paper proposes a phenomenological description of the $2\nu\beta\beta$-decay NMEs, motivated by nuclear theory findings and observed trends in experimental data. The proposed $2\nu\beta\beta$-SEF is
	\begin{equation}
		M^{2\nu{-ph}}= \left(\frac{Z_f}{N_f}\right)^{\alpha}\left(\frac{\Delta_{\rm pn}}{1-\beta_</\beta_>}\right)^\gamma
		\left(T_{f}\right)^{\sigma},
		\label{phnme}
	\end{equation}
	where $Z_f$, $N_f$, and $T_{f}$ are the proton number, neutron number, and the isospin of the final nuclear ground state, respectively. We note that the ground state of the initial (final) even-even nucleus belongs to the isospin multiplet with $T_{i}=(N_i-Z_i)/2$ ($T_{f}=(N_f-Z_f)/2$) and it is the only state of this nucleus belonging to that multiplet for which the isospin projection equals the total isospin. The dependence of $M^{2\nu}_{GT}$ on the value of the isospin was established in the study performed within an exactly solvable model \cite{Stefanik-PRC2015}. The $Z_f/N_f$ and $T_f$ dependence was motivated by observing that the values of $(Z_f/N_f)^{50}T_f^5$, displayed in Figure~\ref{fig:epsart} (and also in Table~\ref{tab:phComp}), show a spread similar to $M^{2\nu-\textrm{exp}}$. The six largest peaks correspond to the NMEs for the $2\nu\beta\beta$-decay of $^{98}$Mo, $^{114}$Cd, $^{104}$Ru, $^{94}$Zr, $^{110}$Pd, and $^{100}$Mo, in this order. While $^{100}$Mo has the largest measured NMEs, theoretical predictions \cite{Nesterenko-TEPJA2022,Suhonen-NPA2011,Pirinen-PRC2015} suggest that the other cases might have larger NMEs than $^{100}$Mo. Unfortunately, these cases have not yet been measured due to their lower $Q$-values \cite{Tretiak-ADNDT2002}. Surprisingly, the simple form $(Z_f/N_f)^{50}~T_f^5$ correctly captures the ordering of $2\nu\beta\beta$-decay NMEs and provides a solid foundation for further refinements.

	The quantities $\beta_<=\min(\beta^{Z},\beta^{Z+2})$ and $\beta_>=\max(\beta^{Z},\beta^{Z+2})$, used in Eq.~(\ref{phnme}), are defined using $\beta^{Z}$ and $\beta^{Z+2}$, the quadrupole deformation parameters of the initial and final nuclei in their ground states, respectively. It is important to emphasize that only prolate shapes for the initial and final nuclear configurations have been considered, as the sign of the quadrupole deformation parameters cannot be inferred from the corresponding electric quadrupole transition probabilities, B(E2). Future experimental data providing insights into the shapes of the nuclei involved in $2\nu\beta\beta$-decay could allow for further refinements to our model. However, for the present study, we have adopted the positive quadrupole deformation parameters listed in the Brookhaven Nuclear Database \cite{NNDC}. The dependence based on the deformation parameters reflects the observation from pn-QRPA \cite{Simkovic-NPA2004,Alvarez-PRC2004}, PHFB \cite{Chandra-TEJPA2005,Chaturvedi-PRC2008}, and NSM \cite{Menendez-NPA2009} calculations that a mismatch in the deformations of initial and final states reduces the NME value for $2\nu\beta\beta$-decay. Additionally, the pairing parameter $\Delta_{\rm pn}$ is a product of experimental pairing gaps \cite{Simkovic-PRC2003},
	\begin{eqnarray}
		\Delta_{\rm pn}
		= \Delta^{\rm Z}_p~ \Delta^{\rm Z}_n~ \Delta^{\rm Z+2}_p~ \Delta^{\rm Z+2}_n.
	\end{eqnarray}
	It reflects the significance of transitions through the lowest $1^+$ states of the intermediate nucleus. The corresponding $\beta$ transition amplitudes are linked to the BCS $u$ and $v$ occupation amplitudes, which can be derived from the gap parameter.
	
	\section{Results and discussions}
	\label{sec:RD}
	The best fitting parameters $\alpha$, $\gamma$, and $\sigma$ have been obtained from the minimization of the chi-squared
	\begin{equation}
		\chi^2 = \sum_{i=1}^N \frac{\left(O_i-P_i\right)^2}{\sigma_i^2},
	\end{equation}  
	where $O_i$ is the experimental NME with uncertainty $\sigma_i$, and $P_i$ is the predicted NME. The sum runs over the direct counter experiments and radiochemical observations. The $2\nu\beta\beta$-decay transitions under the "Fitted" label from Table~\ref{tab:syst} ($N=10$), led to the $2\nu\beta\beta$-SEF with $\alpha=46.94$, $\gamma=0.22$ and $\sigma=4.90$.  Interestingly, when "Fitted" $2\nu\textrm{ECEC}$ cases were added as inputs ($N=12$), but with $T_f \rightarrow T_f + 1$, the same fit parameters were obtained. This might indicate multilevel modeling for both types of transitions, though more half-life measurements for proton-rich nuclei are needed for a clear answer.

	Table~\ref{tab:modelComp} compares the $2\nu\beta\beta$-decay NMEs obtained from the SEF, the SSD hypothesis, and the most recent calculations of different nuclear models. These datasets are compared using $\chi^2/N$ values. The $2\nu\beta\beta$-SEF shows the best agreement with experimental NMEs, generally yielding $\chi^2/N$ values two orders of magnitude smaller than the other calculations. Note that the values for nuclear models in Table~\ref{tab:modelComp} are the calculated $M^{2\nu}_{GT}$, each multiplied by the square of an effective $g_A$ as chosen by the respective papers. We note smaller $\chi^2/N$ values from the SSD hypothesis and the NSM and FSQP calculations, but general conclusions are difficult to draw due to the small number of cases in those datasets.

	\begin{table*}
		\squeezetable
		\caption{The $2\nu\beta\beta$-decay NMEs obtained with the SEF, the ones from the SSD hypothesis, and the most recent calculations from various nuclear structure models. The row labeled "$\chi^2/N$" displays the chi-squared divided by the number of available data for each set. \label{tab:modelComp}} 
		\centering
		\begin{ruledtabular}
			\begin{tabular}{cccccccccccccc}
				& \multicolumn{10}{c}{$M^{2\nu-\textrm{th}}$}&& \multicolumn{2}{c}{$M^{2\nu-{\rm ph}}$} \\
				\cline{2-11} \cline{13-14}
				&QRPA&QRPA &IBM &IBM &IBM &NSM &NSM & PHFB& FSQP&ET&&SEF&SSD \\
				Nucleus&\cite{Pirinen-PRC2015}& \cite{Simkovic-PRC2018}& \cite{Nomura-PRC2024}&\cite{Nomura-PRC2022}& \cite{Barea-PRC2015} &\cite{Patel-NPA2024} &\cite{Caurier-PLB2012}&\cite{Rath-FP2019}& \cite{Ejiri-JPG2017}&\cite{CoelloPerez-PRC2018}&$M^{2\nu-\textrm{exp}}$&&\cite{Kotila-PRC2012}\\
				\hline
				Fitted&&&&&&&&&&&&&\\ 
				\hline
				$^{48}$Ca &--& 0.016 &0.069& 0.024 & 0.045 & -- & 0.039 & -- & -- &-- & $0.0314\pm0.0030$ & 0.022 & --\\
				$^{76}$Ge &--& 0.063 &0.083& 0.018 & 0.085 & -- & 0.097 & -- &0.083 &0.085&$0.0987\pm0.0010$ & 0.105 & --\\
				$^{82}$Se &--& 0.058 &0.072& 0.024 & 0.063 & 0.099 & 0.105 & -- &0.103 & 0.156&$0.0828\pm0.0005$ & 0.081 & --\\
				$^{96}$Zr &--& 0.133 &0.058& 0.054 & 0.034 & -- & -- & 0.092 &0.072 & --&$0.0770\pm0.0040$ & 0.063 & --\\
				$^{100}$Mo &0.105& 0.251 &0.197& 0.157 & 0.045 & -- & -- & 0.164 & 0.154& 0.179&$0.2019\pm0.0016$ & 0.199 & 0.174\\
				$^{116}$Cd & 0.112 & 0.049 &0.089& 0.069 & 0.031 & -- & -- & -- & 0.088 &0.137 &$0.1142\pm0.0027$ & 0.120 & 0.148\\
				$^{130}$Te & 0.057 & 0.053 & 0.035 & 0.010 & 0.038 & 0.027 & 0.036 &0.059 & 0.027 & 0.034&$0.0265\pm0.0003$ & 0.028 & --\\
				$^{136}$Xe & 0.036 & 0.030 & 0.056&0.022 & 0.032 & 0.024 & 0.021 & -- & -- & -- &$0.0174\pm0.0002$ & 0.016 & --\\
				$^{150}$Nd & -- & -- & 0.077 & 0.054 & 0.017 & 0.069 & -- & 0.048 & -- &-- &$0.0527\pm0.0019$ & 0.032 & 0.023\\
				$^{238}$U &--& -- & -- & -- & 0.023 & -- & -- & -- & -- & --&$0.0550\pm0.0110$ & 0.026 & --\\
				\hline
				$\chi^2/N$&5170&2100&2845&2828&1773&470&686&3442&480&4774&--&30&233\\
				\hline
				Predicted&&&&&&&&&&&&&\\ 
				\hline
				$^{110}$Pd & 0.167 & -- &--& 0.022 & 0.041 & -- & -- & -- &0.233& 0.211 &$<2.61$ & 0.147 & --\\
				$^{124}$Sn & 0.023 & -- &--& -- & 0.034 & 0.037 & -- & -- & -- & -- & -- & 0.029 & --\\
				$^{128}$Te & 0.051 & 0.063 &0.022& 0.018 & 0.044 & 0.013 & 0.049 & 0.058 &0.030&0.050 &$0.0366\pm0.0007$ & 0.047 & 0.015\\
				$^{134}$Xe &0.063& -- & -- & -- & -- & -- & -- & -- & -- & --&$<1.25$ & 0.033 & --\\
			\end{tabular}
		\end{ruledtabular}
	\end{table*}   
	
	\begin{table*}
		\squeezetable
		\caption{The $2\nu\beta\beta$-decay NMEs obtained with previous phenomenological models \cite{Ren-PRC2014,Rajan-IJP2018,Pritychenko-NPA2023} compared with those predicted by: (A) $(Z_f/N_f)^{50}T_f^5$, (B) $(Z_f/N_f)^{46.94}T_f^{4.90}$, (C) $(Z_f/N_f)^{46.94}T_f^{4.90}\Delta_{\rm pn}^{0.22}$, (D) $(Z_f/N_f)^{48.44}T_f^{4.95}\left[\Delta_{\rm pn}/(\beta_> -\beta_<)\right]^{0.21}$ and (SEF) $(Z_f/N_f)^{46.94}T_f^{4.90}\left[\Delta_{\rm pn}/(1-\beta_</\beta_>)\right]^{0.22}$. The row labeled "$\chi^2_\nu~(\nu)$" displays the values of reduced chi-squared and, in parenthesis, the DOF for each model. The last column shows the reduced chi-squared when a specific nucleus is involved in the LOOCV for SEF and, in parenthesis, the NME prediction of the excluded case. \label{tab:phComp}} 
		\centering
		\begin{ruledtabular}
			\begin{tabular}{ccccccccccc}
				& \multicolumn{8}{c}{$M^{2\nu-\textrm{ph}}$}&&\\
				\cline{2-9}
				&\multicolumn{3}{c}{Previous models}&\multicolumn{5}{c}{Present models}&&LOOCV for SEF\\
				\cline{2-4} \cline{5-9} \cline{11-11}
				Nucleus&\cite{Ren-PRC2014}&\cite{Rajan-IJP2018}&\cite{Pritychenko-NPA2023}&A&B&C&D&SEF&$M^{2\nu-\textrm{exp}}$&$\chi^2_\nu$ (prediction)\\
				\hline
				Fitted&&&&&&&&&&\\ 
				\hline
				$^{48}$Ca &0.030&0.035& 0.037 &0.008&0.012&0.020&0.023& 0.022& $0.0314\pm0.0030$&48 (0.022)\\
				$^{78}$Ge &0.198&0.123& 0.410 &0.026&0.044&0.068&0.108& 0.105 & $0.0987\pm0.0010$ &41 (0.107)\\
				$^{82}$Se &0.095&0.096& 0.163 &0.015&0.027&0.040&0.080& 0.081 & $0.0828\pm0.0005$& 35 (0.073) \\
				$^{96}$Zr &0.060&0.063& 0.008 &0.027&0.049&0.058&0.068& 0.063 & $0.0770\pm0.0040$ &48 (0.063)\\
				$^{100}$Mo &0.073&0.091& 0.220 &0.045&0.079&0.110&0.197& 0.199 & $0.2019\pm0.0016$ &47 (0.183)\\	
				$^{116}$Cd &0.071&0.093& 0.121 &0.031&0.058&0.081&0.129& 0.120 & $0.1142\pm0.0027$&49 (0.121) \\		
				$^{130}$Te &0.077&0.068& 0.032 &0.006&0.014&0.016&0.029& 0.028 & $0.0265\pm0.0003$ &32 (0.030)\\
				$^{136}$Xe &0.030&0.055& 0.017 &0.004&0.010&0.012&0.016& 0.016 & $0.0174\pm0.0002$ &40 (0.015)\\
				$^{150}$Nd &0.030&0.061& 0.203 &0.009&0.021&0.025&0.027& 0.032 & $0.0527\pm0.0019$ &29 (0.031) \\
				$^{238}$U &0.024&--& 0.046 &0.005&0.014&0.010&0.020& 0.026 & $0.0550\pm0.0110$  &48 (0.026) \\
				\hline
				$\chi^2_\nu~(\nu)$&5651~(9)&24981~(2)&17632~(7)&5450~(8)&3171~(8)&2017~(7)&67~(7)&43~(7)&&\\
				\hline
				Predicted&&&&&&&&&&\\ 
				\hline
				$^{110}$Pd &0.135&0.160& 0.270 &0.047&0.084&0.115&0.143& 0.147 & $<2.61$  &   \\
				$^{124}$Sn &0.098&--& 0.019 &0.009&0.018&0.025&0.029& 0.029 & --    & \\
				$^{128}$Te &0.067&--& 0.042 &0.014&0.030&0.037&0.045& 0.047 & $0.0366\pm0.0007$   &  \\
				$^{134}$Xe &--&--& 0.086 &0.010&0.022&0.025&0.033& 0.033 & $<1.25$   &  \\
			\end{tabular}
		\end{ruledtabular}
	\end{table*}
	
	We also compare the $2\nu\beta\beta$-SEF with earlier empirical models and evaluate its design, stability, and predictive power. Table \ref{tab:phComp} presents the NMEs from present and previous phenomenological models \cite{Ren-PRC2014,Rajan-IJP2018,Pritychenko-NPA2023} alongside the experimental values. To assess the goodness of fit, we computed the reduced chi-squared $\chi^2_{\nu}=\chi^2/\nu$, where $\nu$ is the number of degrees of freedom (DOF). We note considerably larger $\chi^2_{\nu}$ values for the previous models compared with the $2\nu\beta\beta$-SEF proposed in this paper. Notably, the model from \cite{Rajan-IJP2018} might risk overfitting, with only two DOF for such a small experimental dataset.

	To better understand the foundation of the $2\nu\beta\beta$-SEF, the precision of $(Z_f/N_f)^{50}T_f^5$ form has been progressively improved by adjusting fit parameters and increasing model complexity. The obtained analytical forms are provided in the caption of Table~\ref{tab:phComp}. Surprisingly, the $\chi^2_\nu$ value of model A is comparable to the best previous model from \cite{Ren-PRC2014}. The results from model B demonstrate that a relatively modest reduction in the powers governing $Z_f/N_f$ and $T_f$ leads to a large step toward the experimental NMEs. The subsequent models, which incorporate pairing parameters and quadrupole deformation, reveal that the inclusion of deformation overlap represents the most significant refinement in the form that initially considered only $Z_f/N_f$ and $T_f$ dependence. We also note that the inclusion of the deformation overlap in the form $(1-\beta_</\beta_>)$ is preferable over $(\beta_> - \beta_<)$, given the current experimental data.

	We applied leave-one-out cross-validation (LOOCV) to assess the robustness of the $2\nu\beta\beta$-SEF in describing the experimental data and making predictions. LOOCV involves systematically excluding one data point at a time from the fitting procedure and evaluating the performance of the $2\nu\beta\beta$-SEF on the remaining data. The values of $\chi^2_{\nu}$ for the SEF with each exclusion are presented in the last column of Table~\ref{tab:phComp}. The NME prediction for the excluded case is presented in parentheses. Excluding a specific case scarcely influences the predictions. Moreover, the small variations in $\chi^2_{\nu}$ values confirm the SEF's stability and predictive power.

	At first glance, the prediction for the $2\nu\beta\beta$-decay NME for $^{128}$Te appears to disagree with experimental data. However, it is important to note that its NME experimental value is derived based on geochemical measurements of ancient tellurium ores \cite{Bernatowicz-PRC1993}. These measurements should be treated with caution, as unknown processes during the formation of these ores may have influenced the daughter isotope production. For example, it has been suggested in \cite{Dirac-N1937,Barabash-JETPL1998} that these results could be affected by a potential time variation in the weak interaction strength. These considerations may also impact measurements of $^{130}$Ba and $^{132}$Ba. Due to these uncertainties, we have excluded all geochemical data from the fit. The radiochemical measurement of $^{238}$U was included, as its larger uncertainty reduces the risk of skewing the fit parameters (see LOOCV with $^{238}$U from Table~\ref{tab:phComp}).
	
	The $2\nu\beta\beta$-SEF predicts quite different NMEs for pairs of isotopes differing by two neutrons, a feature also observed in modern nuclear structure models from Table~\ref{tab:modelComp}. For instance, the SEF predicts NME ratios of around $2$ for the pairs $^{128,130}$Te and $^{134,136}$Xe, which is quite different from Pontecorvo's assumption that such pairs must have nearly equal NMEs \cite{Pontecorvo-PLB1968}. Similar pairs include $^{98,100}$Mo, $^{114,116}$Cd, and $^{94,96}$Zr, although the small $Q$-values of the two-neutron-less isotopes limit near-future measurements for these cases. Fortunately, the $2\nu\beta\beta$-decay of $^{134}$Xe might soon be observed, as it is an unavoidable source of background in Dark Matter searches \cite{XENONnT-JCAP2020,LUX-ZEPLIN-PRD2020,DARWIN-JPG2022,PandaX-4T-arXiv2024}. We also note the SEF's optimistic half-life predictions, ranging between $10^{20}-10^{21}$ years, for the $2\nu\beta\beta$-decay of $^{110}$Pd and $^{124}$Sn, as well as for the $2\nu$ECEC of $^{106}$Cd. These nuclei, with sufficiently large $Q$-values, are already being considered as candidates for near-future measurements \cite{Fink-PRL2012,Nanal-EJPWC2014,Chkvorets-arViv2017,Rukhadze-JPCS2021,Belli-P2021}. Future direct counter experiments could expand on and confirm the reliability of the SEF, which, using only a few parameters, has successfully accounted for the large spread of the measured $2\nu\beta\beta$-decay NMEs.\\

	\section{Conclusions}
	\label{sec:Conclusions}
	
	In this paper, we proposed a semi-empirical description of the NMEs for two-neutrino double beta decay. Inspired by the findings of nuclear many-body methods and trends in experimental data, the $2\nu\beta\beta$-SEF depends on the proton and neutron numbers, pairing, isospin, and deformation properties of the initial and final nuclei. We also found an indication of multilevel modeling for both $2\nu\beta\beta$-decay and $2\nu\textrm{ECEC}$ process, but more measurements for proton-rich nuclei are required for a definitive answer. A comprehensive comparison with previous phenomenological and nuclear models revealed that the proposed SEF yields the best agreement with the experimental data, explaining the large spread of measured $2\nu\beta\beta$-decay NMEs. The stability of the SEF was confirmed by the LOOCV and reasons to exclude geochemical measurements as inputs were discussed. Finally, we noted that the SEF's predictions of the NMEs for $2\nu\beta\beta$ nuclear systems differing by two neutrons, are significantly different (by about a factor $2$), in contrast with the previous assumptions. Its correctness can be checked by measuring additional $2\nu\beta\beta$ transitions.

	\section*{Acknowledgments}
	O.N. acknowledges support from the Romanian Ministry of Research, Innovation, and Digitalization through Project No. PN 23 08 641 04 04/2023. F.\v{S}. acknowledges support from the Slovak Research and Development Agency under Contract No. APVV-22-0413, VEGA Grant Agency of the Slovak Republic under Contract No. 1/0618/24 and by the Czech Science Foundation (GA\v{C}R), project No. 24-10180S.

	\bibliographystyle{apsrev4-2}
	\bibliography{bibliography}

\end{document}